\newif\iffigs\figsfalse
\figstrue

\documentstyle[psfig,11pt,a4]{article}
\iffigs
\input epsf
\else
\fi 

\begin{document}



\newcounter{subequation}[equation]

\makeatletter
\expandafter\let\expandafter\reset@font\csname reset@font\endcsname
\newenvironment{subeqnarray}
{\def\@eqnnum\stepcounter##1{\stepcounter{subequation}{\reset@font\rm
  (\theequation\alph{subequation})}}\eqnarray}%
{\endeqnarray\stepcounter{equation}}
\makeatother


\newcommand{\ga}{\alpha}
\newcommand{\gb}{\beta}
\newcommand{\gc}{\gamma}
\newcommand{\gcp}{\gamma^\prime}
\newcommand{\gd}{\delta}
\newcommand{\gep}{\epsilon}
\newcommand{\gl}{\lambda}
\newcommand{\gL}{\Lambda}
\newcommand{\gk}{\kappa}
\newcommand{\go}{\omega}
\newcommand{\gp}{\phi}
\newcommand{\gs}{\sigma}
\newcommand{\gt}{\theta}
\newcommand{\gC}{\Gamma}
\newcommand{\gD}{\Delta}
\newcommand{\gO}{\Omega}
\newcommand{\gT}{\Theta}
\newcommand{\gvp}{\varphi}


\newcommand{\be}{\begin{equation}}
\newcommand{\ee}{\end{equation}}
\newcommand{\ba}{\begin{array}}
\newcommand{\ea}{\end{array}}
\newcommand{\bea}{\begin{eqnarray}}
\newcommand{\eea}{\end{eqnarray}}
\newcommand{\bes}{\begin{eqnarray*}}
\newcommand{\ees}{\end{eqnarray*}}
\newcommand{\bsea}{\begin{subeqnarray}}
\newcommand{\esea}{\end{subeqnarray}}
\newcommand{\lra}{\longrightarrow}
\newcommand{\lms}{\longmapsto}
\newcommand{\ra}{\rightarrow}
\newcommand{\pa}{\partial}


\newcommand{\N}{\mbox{I\hspace{-.4ex}N}}
\newcommand{\C}{\mbox{$\,${\sf I}\hspace{-1.2ex}{\bf C}}}
\newcommand{\Cs}{\mbox{$\,${\sf I}\hspace{-1.2ex}C}}
\newcommand{\Z}{\mbox{{\sf Z}\hspace{-1ex}{\sf Z}}}
\newcommand{\R}{\mbox{\rm I\hspace{-.4ex}R}}
\newcommand{\1}{\mbox{1\hspace{-.6ex}1}}


\makeatletter
\newcommand{\Tr}{\mathop{\operator@font Tr}\nolimits}
\newcommand{\me}{\mathop{\operator@font e}\nolimits}
\newcommand{\th}{\mathop{\operator@font th}\nolimits}
\newcommand{\ch}{\mathop{\operator@font ch}\nolimits}
\newcommand{\sh}{\mathop{\operator@font sh}\nolimits}
\newcommand{\co}{\mathop{\operator@font c}\nolimits}
\newcommand{\si}{\mathop{\operator@font s}\nolimits}
\makeatother

\newcommand{\la}{\wi{\Tr J^2}}
\newcommand{\DI}[1]{\mbox{$\displaystyle{#1}$}}
\newcommand{\wi}[1]{\widehat{#1}}



\thispagestyle{empty}

\hbox to \hsize{%
\vtop{\hbox{      }\hbox{     }} \hfill
\vtop{\hbox{DAMTP-R-98-26}}}

\vspace*{1cm}

\bigskip\bigskip\begin{center}
{\bf \Huge{'t Hooft's Polygon Approach Hyperbolically Revisited}}
\end{center}  \vskip 1.0truecm
\centerline{{\bf Helia R. Hollmann\footnote{e-mail: H.Hollmann@damtp.cam.ac.uk} 
and Ruth M. Williams\footnote{e-mail: R.M.Williams@damtp.cam.ac.uk}}}
\centerline{Department of Applied Mathematics and Theoretical Physics,}
\centerline{Silver Street, Cambridge CB3 9EW, England}
\vskip 2cm
\bigskip \nopagebreak \begin{abstract}
\noindent
The initial data in the polygon approach to (2+1)D gravity coupled 
to point particles are constrained by the vertex equations and the 
particle equations. 
We establish the hyperbolic nature of the vertex equations and derive 
some consequences. In particular we are able to identify the 
hyperbolic group of motions as discrete analogues of the diffeomorphisms 
in the continuum theory. 
We show that particles can be included ``hyperbolically'' as well, 
but they spoil the gauge invariance. 
Finally we derive consistent sets of initial data.

\end{abstract}

\newpage\setcounter{page}1

\section{Introduction}

As a response to a paper by Gott \cite{Got91} establishing  
acausal features in (2+1)D gravity coupled to point particles,  
't Hooft invented his polygon approach \cite{'tH92}. 
He succeeded in showing that even in the case of two particles of 
equal mass and rapidity approaching each other nearly head--on, 
the space--time would have to undergo a big crunch before a 
closed timelike curve could be formed. 

His idea to split the space--time into a direct product of a cosmological 
time and a Cauchy surface tessellated by entirely flat polygons 
turned out to be a particularly useful approach to describe particles 
in a (2+1)D gravity theory. In the last couple of years toy models 
have been constructed \cite{Wel97}, \cite{'tH931}, issues of topology 
have been addressed \cite{FraGua96}, \cite{Wel97}, particle decay 
and space--time kinematics have been investigated \cite{FraGua95},
and inspired by the polygon approach quantized models of (2+1)D 
space--time have been invented \cite{'tH932}, \cite{'tH96}, 
\cite{Wae94}. 

As a (2+1)D theory of gravity has very peculiar features we do 
not claim that we study the theory in order to get any particular  
insight into the (3+1)D equivalent. 
Due to the fact that in (2+1)D the Riemannian curvature tensor 
is algebraically dependent on the energy momentum tensor the 
space--time is locally flat outside the sources. In the case of 
pure gravity topology might introduce a certain degree of 
nontriviality. There an open neighbourhood isometric to Minkowski 
space cannot be extended to include all of the manifold because 
of nontrivial gluing homomorphisms. If we include particles,
we are confronted with the fact that the theory has no 
Newtonian limit. Test particles in a Newtonian theory experience 
a logarithmic gravitational potential outside the matter sources 
and hence accelerate. In an alternative classical or a quantum 
theory of gravity  one would certainly like to test the results 
against the classical theory or to establish some kind of classical 
limit. Here this is not possible. But let us follow 't Hooft's ideas 
more closely. His approach -- roughly speaking -- is an  
implementation of the local flatness of space--time in the 
pure gravity scenario supplemented by the additional conelike  
structure introduced by particles. Again, in (3+1)D gravity a 
particle would not simply cut a cone out of space--time. 

What are our motivations? We would like to address two issues mainly. 
First of all (2+1)D gravity coupled to point particles is interesting 
in its own right. In particular it is a nontrivial task to follow the time 
evolution of a (2+1)D space--time with particles. The system has finitely 
many degrees of freedom but -- as is well known -- even for  a four 
particle system the analytical solution is not known in classical 
mechanics. Therefore we would like to perform a computer simulation 
of particle universes tesselated by polygons. 
't Hooft himself \cite{'tH931} wrote a computer program to evolve 
point particles in (2+1)D. Unfortunately he implemented the code 
on a very small (40 kbytes memory) and old home computer. 
Somehow the code does not seem to be transferable to a modern machine.   
His big bang and big crunch hypothesis as well as other results 
\cite{'tH931} are based on a 1--polygon tesselated Cauchy surface 
with either $S^2$ or $S^1 \times S^1$ topology. 
We would like to test them against more complex initial configurations. 
Furthermore there are several approaches to (2+1)D gravity with 
and without particles, under discussion. We aim at comparing 
several of them, starting with the polygon approach and Regge calculus. 
This paper is a step in both directions. 

We start by reviewing the basic setup for the polygon approach 
to make the paper sufficiently self-contained. 
Then we first investigate the pure gravity sector the initial 
data of which is described by the so called vertex equations. 
Already 't Hooft \cite{'tH933} noticed that these equations are 
related to trigonometric properties of triangles in a hyperbolic 
space. We establish this relation in detail proving that the vertex 
relations can be reduced to the hyperbolic law of sines and the first 
and second law of hyperbolic cosines.  The laws of sines and cosines 
for the edges can be derived  from the law of cosines for the angles, 
leading to the result that only one 
equation out of the set of vertex relations is independent. We 
decided to take the hyperbolic point  of view seriously. This leads 
to interesting consequences. 
A close investigation of the vertex relations then yields  the 
result that it is the freedom to choose frames 
which make the hyperbolic configurations possible. 
As a next step we establish the hyperbolic nature of a particle vertex. 
The particle equations turn out to be the trigonometric relations 
in a rectangular hyperbolic triangle. Of course, here the hyperbolic 
geometry cannot disappear by a gauge transformation. 
In the last chapter we summarize some of 
the consistent initial configurations  we have found.      
   
\section{Pure Gravity Sector}

Let the (2+1)D space--time be foliated by two--dimensional 
cauchy surfaces. We use the feature that vacuum space--time 
is entirely flat and tessellate the spatial part of it by polygonal tiles. 
To each of these pieces of $\R^2$ we attach a Lorentz frame. 
We impose two further constraints. First, let us assume that time 
runs equally fast for all the polygons, i.e. they have a common 
clock. This corresponds to partially fixing the gauge. 
Secondly we consider 3--valent vertices only. This does not impose 
restrictions on the physics of our system as all other 
configurations can be decomposed into configurations with 
3--valent vertices and edges with zero length connecting them. 

The consequences of these assumptions are proven in \cite{'tH92} (we 
call them ``initial consequences'' in the following):  
The lengths of adjacent edges measured in the reference frame 
of each polygon are equal. The velocity of the edges turns out to be 
orthogonal to the orientation of the edges viewed in the rest frame 
of one polygon as well as with respect to the other polygon's frame. 
They are equal up to signs. 

\clearpage
\iffigs
\begin{figure}[bt]
\begin{center}
\begin{minipage}[t]{11cm}
  \epsfxsize=10cm\epsfbox{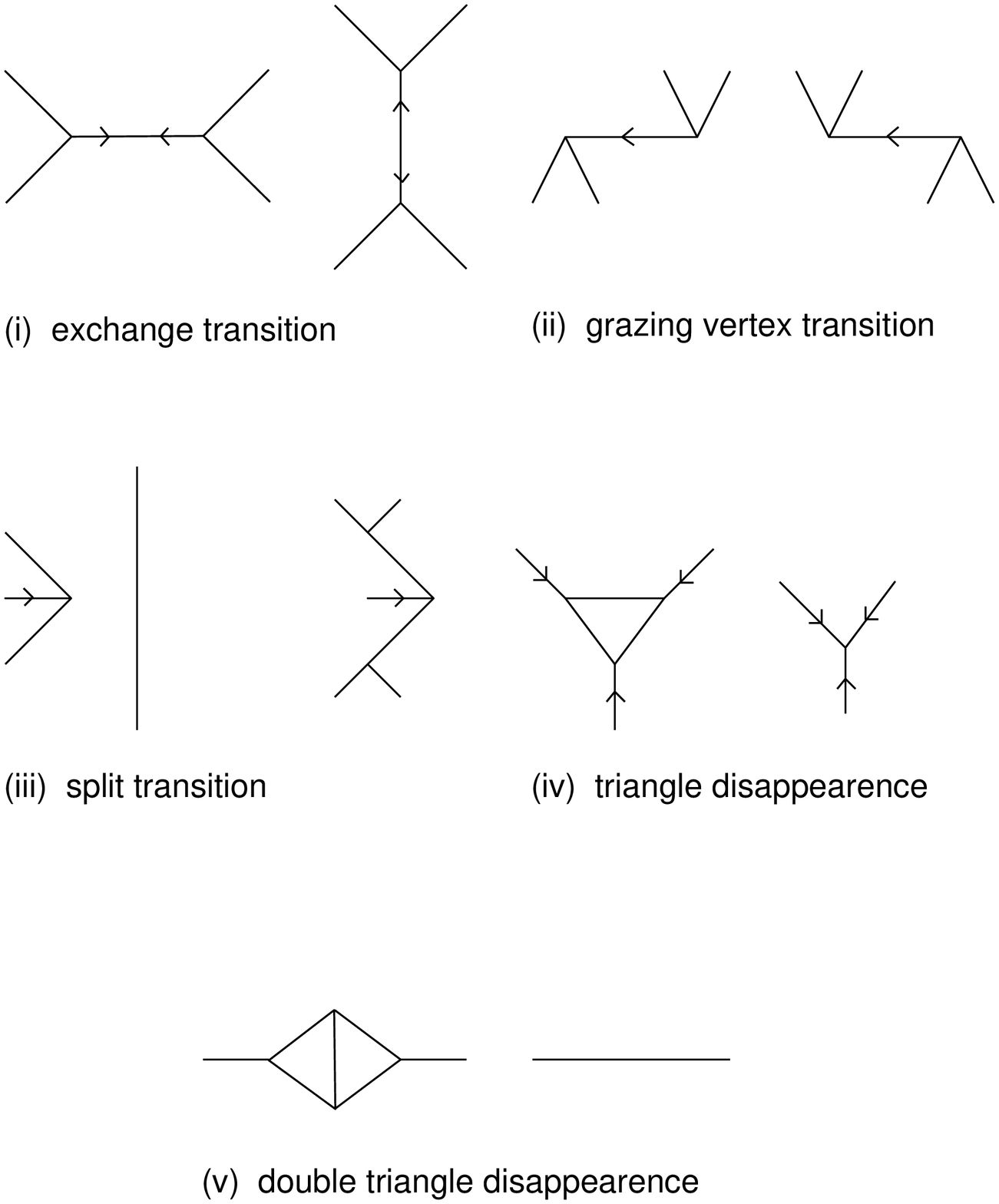}
  \small{{\bf figure 1:} pure gravity transitions}
\end{minipage}
\end{center}
\end{figure}
\fi

For reasons of consistency these features have to be valid at 
every instant of time leading to so called transition rules (see 
{\bf figure 1}) which   have been classified by 't Hooft \cite{'tH931}. 
The transition rules have to be taken into account whenever the 
polygonal configuration changes. 

In the vacuum case this is happening 
when one or more edges shrink to zero or a vertex hits an edge leading 
to an intermediate configuration of several 3--valent vertices connected 
by edges of zero length. 
As the edges ``carry'' a velocity their lengths will change in time. 
There is a growth rate associated with every vertex. 
That results in time evolution equations for the edge lengths $L_1$ 
and $L_2$ of the straight lines between the vertices $A$ and $B$ or 
$A$ and $C$, respectively. The rate of change of the $L_i$'s in time 
is given by 
\[
   \dot{L_1} = g_{A,1} + g_{B,1}, \qquad 
   \dot{L_2} = g_{A,2} + g_{C,2}. 
\]
$g_{A,1}, g_{A,2}, g_{B,1}, g_{C,2}$ are the growth rates associated 
with the vertices $A, B, C$ and the edge lengths $L_1$ and $L_2$, 
respectively. 
Let us focus on the vertex $A$ as shown in {\bf figure 2}. The growth 
rates $g_{A,1}$ and $g_{A,2}$ are read off to be 
\[
  g_{A,1} = \frac{v_2 + v_1 \cos \ga_3}{\sin \ga_3}, \qquad
  g_{A,2} = \frac{v_1 + v_2 \cos \ga_3}{\sin \ga_3}.
\]
$g_{B,1}$ and $g_{C,2}$ are calculated in an analogous manner. 

\clearpage
\iffigs
\begin{figure}[bt]
\begin{center}
\begin{minipage}[t]{8cm}
  \epsfxsize=7cm\epsfbox{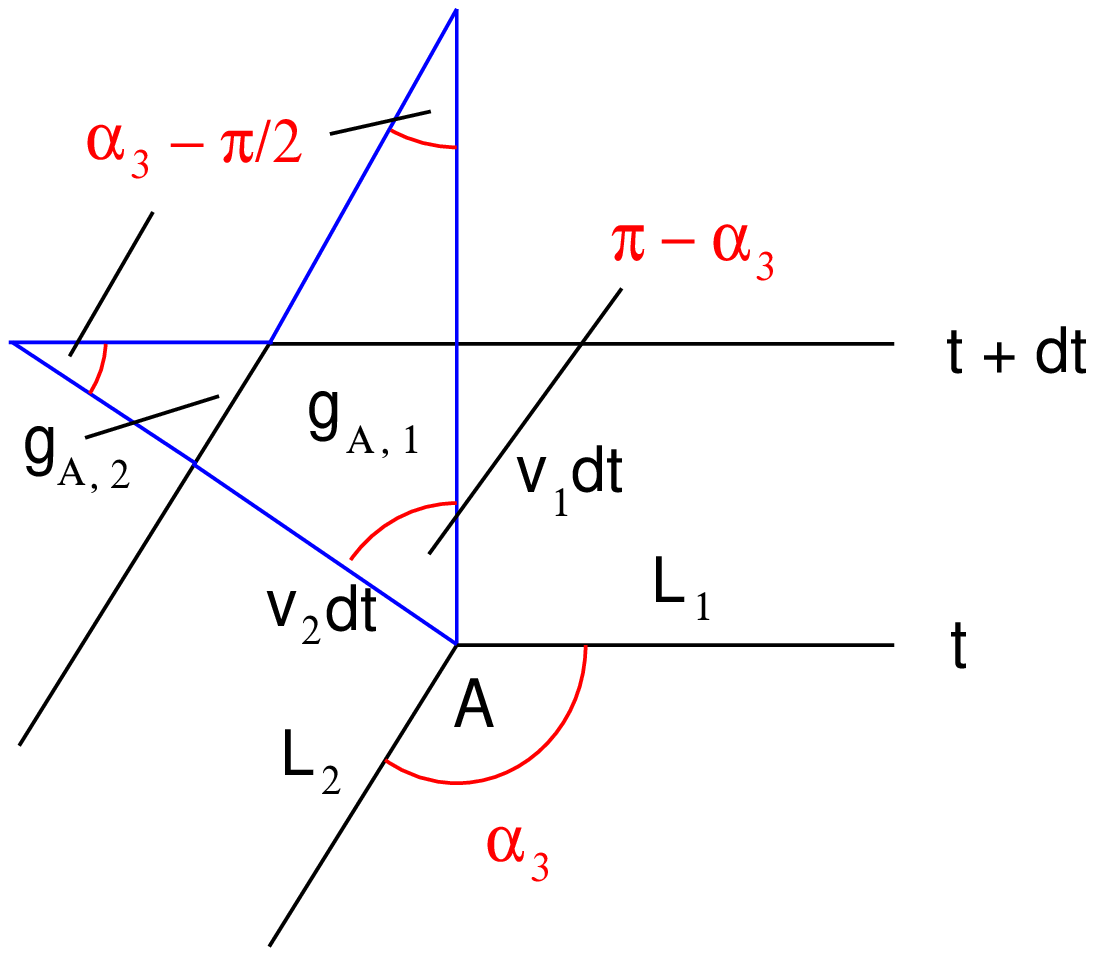}
  \small{{\bf figure 2:} equations of motion for gravity vertices}
\end{minipage}
\end{center}
\end{figure}
\fi

From an algorithmic point of view evolution of a polygon universe 
in pure gravity means the following: we start with a proper set 
of initial data, that 
is a number of polygons, the coordinates of their vertices (and therefore 
the length of the edges joining them) and the velocities with 
which they move. We then let the configuration evolve linearly 
until a transition takes place. 
The transition rules reshuffle the set of initial data 
into another one, which has in general a different number of polygons, 
edges and vertices still obeying the initial conditions. 
The initial data cannot be chosen freely. 

At the edges, the frames of 
adjacent polygons are related by a Lorentz transformation. 
Let us circle around a vertex, which is a meeting point of 
three polygons and therefore of three reference frames. 
To circle around a vertex means to undergo a sequence of Lorentz boosts 
in the $tx$--plane with boost parameters $2 \: \eta_i$ and subsequent 
rotations with angles $\ga_i$ in the $xy$--plane. Assuming a 
counterclock--wise orientation they combine to the following identity 
due to the flatness of space--time 
\[
  L_2 \: R_3 \: L_1 \: = \: R_1^{-1} \: L_3^{-1} \: R_2^{-1}, 
\]
with 
\[
  R_i = \left(
        \ba{ccc} 
        1 & 0 & 0 \\
        0 & \cos \ga_i & \sin \ga_i \\
        0 & -\sin \ga_i & \cos \ga_i
        \ea 
        \right) \qquad \mbox{and} \qquad
  L_i = \left(
        \ba{ccc} 
        \cosh 2\eta_i & \sinh 2\eta_i & 0 \\
        \sinh 2\eta_i & \cosh 2\eta_i & 0  \\
        0 & 0 & 1
        \ea 
        \right). 
\]
In detail the equations are
\bsea
  \ch_3 &=& \ch_1 \ch_2 \: + \: \co_3 \sh_1 \sh_2 \\
  - \co_2 \sh_3 &=& \ch_2 \sh_1 \: + \: \co_3 \ch_1 \sh_2 \\
  \si_2 \sh_3 &=& \sh_2 \si_3 \\
  - \co_1 \sh_3 &=& \ch_1 \sh_2 \: + \: \co_3 \ch_2 \sh_1 \\
  \co_1 \co_2 \ch_3 \: - \: \si_1 \si_2 
      &=& \sh_1 \sh_2 \: + \: \co_3 \ch_1 \ch_2 \\
  - \co_1 \si_2 \ch_3 \: - \: \si_1 \co_2 &=& \si_3 \ch_2 \\
  - \si_1 \sh_3 &=& - \si_3 \sh_1 \\
  \si_1 \co_2 \sh_3 \: + \: \co_1 \si_2 &=& - \si_3 \ch_1 \\
  - \si_1 \si_2 \ch_3 \: + \: \co_1 \co_2 &=& \co_3
\esea
$\si_i, \co_i$ denote $\sin \ga_i, \cos \ga_i$ and  $\sh_i, \ch_i$ are 
$\sinh 2\eta_i, \cosh 2\eta_i$. 
By taking into account the permutation of indices or by fixing one 
index and interchanging the other two\footnote{that is, we change 
the orientation of the rotations}, we find that (1a) is equivalent 
to (A.5), (1c) and (1g) correspond to (A.2), (1f) and (1h) are (A.3) 
and (1i) leads to (A.4). (A.2) -- (A.5) refer to 't Hooft's notation 
in \cite{'tH92}. For the convenience of the reader we include his equations 
here. 
\bea
  & & \si_1 \: : \: \si_2 \: : \: \si_3 
       = \sh_1 \: : \: \sh_2 \: : \: \sh_3 
                       \nonumber \hspace{5cm} \mbox{(A.2)} \\
  & & \ch_2 \si_3 \: + \: \si_1 \co_2 \: + \: \co_1 \si_2 \ch_3 \: = \: 0 
       \nonumber \hspace{5.1cm} \mbox{(A.3)}\\
  & & \co_1 \: = \: \co_2 \co_3 \: - \:  \ch_1 \si_2 \si_3 
       \nonumber \hspace{6.4cm} \mbox{(A.4)}\\
  & & \ch_1 \: = \: \ch_2 \ch_3 \: + \: \sh_2 \sh_3 \co_1 
       \nonumber \hspace{5.6cm} \mbox{(A.5)}\\
  & & \cot_2 \: = \: - \cot_1 \ch_3 \: - \: \frac{\coth_2 \sh_3}{\si_1} 
   \nonumber \hspace{5cm} \mbox{(A.6)} 
\eea
(1b) and (1d) are equivalent. They arise using the 
identity $\si_i = c \: \sh_i$\footnote{which is given by the vertex 
equations (1c) and (1g), respectively: $c = \si_j/\sh_j$} with cyclic 
permutations from (A.3). 
That equation (1e) is not independent can be seen in the following 
manner: we start with 
\[
  \si_1 \si_2 \sh_3^2 \: = \: \si_3^2 \sh_1 \sh_2.        
\]
Equivalently we get 
\[
  \ch_3 \co_3 \: + \: \ch_3^2 \si_1 \si_2 \: - \: \si_1 \si_2 
    \: = \: \ch_3 \co_3 \: - \: \co_3^2 \sh_1 \sh_2 \: + \: \sh_1 \sh_2.
\]
With (1i) and (1a) we derive
\[
  \ch_3 \co_1 \co_2 \: - \: \si_1 \si_2 
      \: = \: \co_3 \ch_1 \ch_2 \: + \: \sh_1 \sh_2, 
\]
which is equation (1e). 

As the next step we show that at most only two of 't Hooft's 
equations, (A.2) and (A.5), are independent. 
We already got the result that (1d) is a 
consequence of (A.2) and (A.3), and (1e) follows with (A.2), (A.3) 
and (A.5). Therefore it is sufficient to show that (A.3) and (A.4) 
can be derived using (A.2) and (A.5). 
Indeed, if we start with a cyclic permutation of (A.5) we find  
\[
  \ch_2 \: = \: \ch_3 \ch_1 \: + \: \co_2 \sh_1 \sh_3 
        \: = \: \ch_3 ( \ch_2 \ch_3 \: + \: \co_1 \sh_2 \sh_3) 
               \: + \: \co_2 \sh_1 \sh_3. 
\] 
This is equivalent to 
\[
  0 \: = \: \sh_3 \: ( \sh_3 \ch_2 \: + \: \co_1 \sh_2 \ch_3 
     \: + \: \co_2 \sh_1).
\]
(A.3) follows by using (A.2). 
On the other hand, if we begin our calculations with (A.3)  
with indices 2 and 3 interchanged and substitute this into (A.3) 
we derive  (A.4) with indices 1 and 2 interchanged. 

In addition to the equations (A.2) -- (A.5), 't Hooft uses the 
equation (A.6) which is a combination of (A.2) and (A.3). 

To summarize these results let us now replace the angle $\ga$ 
by $\pi - \ga$. Then the equations (A.4) and (A.5) become 
\bea
  \qquad \qquad \si_2 \si_3 \ch_1 &=& \co_2 \co_3 \: + \: \co_1  
       \hspace{6cm} \mbox{(A.4$^\prime$)} \nonumber \\
  \qquad \qquad \ch_1 &=& \ch_2 \ch_3 \: - \: \co_1 \sh_2 \sh_3    
       \hspace{4.4cm}  \mbox{(A.5$^\prime$)} \nonumber
\eea
We identify \cite{Gre80} (A.5$^\prime$) with the first law and 
(A.4$^\prime$) with the second law of hyperbolic cosines. 
(A.2) is the law of hyperbolic sines. 
But we succeed in showing even more:  
(A.2) and (A.5) are a consequence of (A.4). That is due to the fact that 
in hyperbolic geometry the cosine and sine formulas are not independent 
of each other. The laws of hyperbolic sines and cosines for the edges 
can be derived from the law of cosines for the angles. 

Let us first prove that (A.2) follows with (A.4). 
Let $\si_2 \si_3$ be unequal to zero. The case $\si_2 \si_3$ equal to 
zero is governed by the so called trivial and quasistatic vertices, 
which we discuss below.  
We can write 
\[
  \sh_1^2 \: = \: \ch_1^2 \: -\: 1 
   \: = \: \frac{(\co_2 \co_3  - \co_1)^2}{\si_2^2 \si_3^2} 
    \: - \: \frac{(1-\co_2^2) (1 - \co_3^2)}{\si_2^2 \si_3^2}. 
\]
Consequently the following equation holds by a cyclic permutation
\[
  \si_2^2 \si_3^2 \sh_1^2 
   \: = \: \co_1^2 \: + \: \co_2^2 \: + \: \co_3^2 \: -\:  1 
        \: -\:  2 \co_1 \co_2 \co_3  
   \: = \: \si_1^2 \si_3^2 \sh_2^2 
\]
and therefore  we find 
\[
  \frac{\sh_1^2}{\sh_2^2} \: = \: \frac{\si_1^2}{\si_2^2} 
    \qquad \mbox{and} \qquad
  \frac{\sh_1}{\sh_2} \: = \: \pm \frac{\si_1}{\si_2}. 
\]
Next we show that (A.5) follows from (A.2) and (A.4). We assume  
that $\si_1 \si_2 \si_3$ is unequal to zero. 
\bea
  \ch_1 \: - \: \ch_2 \ch_3 &=& \frac{\co_2 \co_3 - \co_1}{\si_2 \si_3} 
    \: - \: \frac{(\co_1 \co_3 - \co_2)(\co_1 \co_2 - \co_3)}{
        \si_1^2 \si_2 \si_3} 
   \nonumber \\
  &=& \co_1 \left( 
    \frac{-1 \:-\: 2 \co_1 \co_2 \co_3 + \co_1^2 + \co_2^2 + \co_3^2}{
      \si_1^2 \si_2 \si_3} 
            \right) \nonumber \\
  &=& \frac{\co_1 \si_2^2 \si_3^2 \sh_1^2}{ \si_1^2 \si_2 \si_3} 
   \: = \: \frac{ \co_1 \si_2 \si_3 \sh_1^2}{\si_1^2} 
   \: = \: \co_1 \sh_2 \sh_3 \nonumber 
\eea
and therefore
\[
  \ch_1 \: - \: \ch_2 \ch_3 \: = \: \co_1 \sh_2 \sh_3.
\]
We have proved that the vertex equations introduced by 't Hooft are 
not an independent set of equations. They can be reduced to one equation. 
They are nothing but the defining relation of a triangle in 
hyperbolic space. The inner angles of the triangle are $\pi - \ga_i$, 
where  $\ga_i$ are the angles between the edges meeting at a 3--valent vertex. 
The edges of the triangles are given by the boost parameters $2 \eta_i$. 
Whereas in hyperbolic geometry the edge length of the triangles are 
assumed to be greater than zero, that is no longer the case here. 
The $\eta_i$'s can be positive or negative. 

The axioms of hyperbolic geometry immediately lead to the consequence 
that the sum of the inner angles in a triangle is less than $\pi$.
Triangles which are similar are already congruent. 
If the configuration is ``truly hyperbolic'', that is 
$\sum (\pi - \ga_i) < \pi$ then we conclude that 
$\sum \ga_i > 2 \pi$ and the geometry of our polygons is elliptic. 
If the angles do not add up to $2 \pi$, we get the result that -- still 
imposing a flat configuration -- the edges are not represented by 
straight lines. For hyperbolic configurations we use for purposes 
of illustration the Poincare model on the open unit  
disk. In this manner the straight lines forming a Euclidean triangle 
get replaced by segments of circles (see {\bf figure 3a}). The circles 
have the property that they intersect the unit circle orthogonally. 
For elliptic geometry it is still possible to use segments of circles 
but the pieces which build up a triangle or a more general polygon 
are concave rather than convex (see {\bf figure 3b}). 

One particular solution of the vertex equations, which exists for a 
given deficit angle $\gb, 0 \le \gb < \pi$, involves equilateral 
hyperbolic triangles. The angles $\ga_1 = \ga_2 = \ga_3 = \ga$ and the 
boosts $\eta_1 = \eta_2 = \eta_3 = \eta$ fulfil the vertex equations. 
To every hyperbolic triangle there corresponds a vertex  (see 
{\bf figure 4}). As mentioned above the angles of the hyperbolic triangle 
are $\pi - \ga_i, 1 \le i \le 3$, and $ \sum ( \pi - \ga_i) < \pi$. 
Consequently the angles at a vertex add up to 
$\sum \ga_i > 2 \pi$. In particular, they do not add up to $2 \pi$ and the 
edges cannot be represented by straight lines. The overlap of the 
angles is given by the deficit angle. The edges are represented by 
segments of a circle, which carry a boost vector.     

\iffigs
\begin{figure}[bt]
\begin{minipage}[t]{6cm}
  \epsfxsize=5.5cm\epsfbox{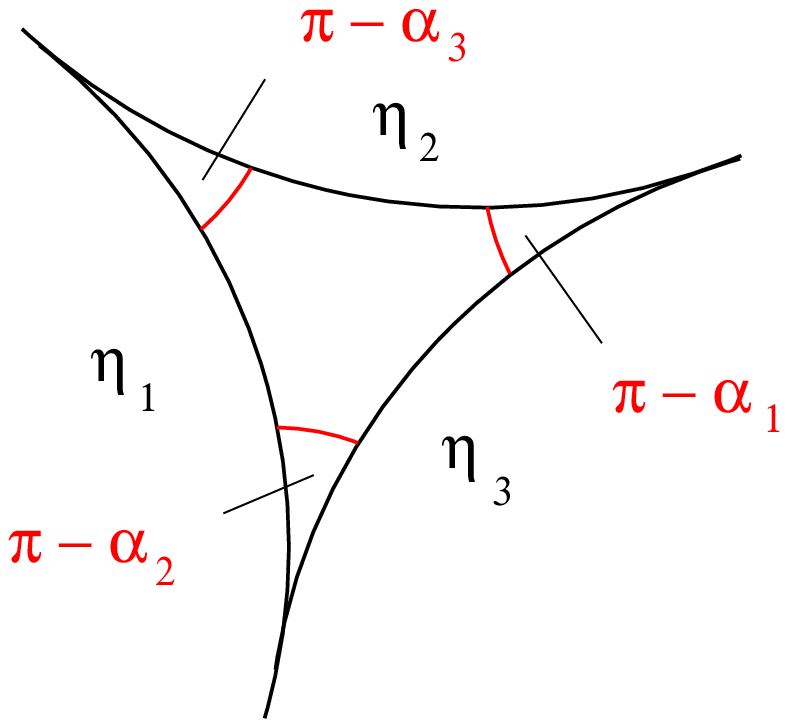}
  \small{{\bf figure 3a:} The hyperbolic triangle.}
\end{minipage}
\hfill
\begin{minipage}[t]{6cm}
  \epsfxsize=5.5cm\epsfbox{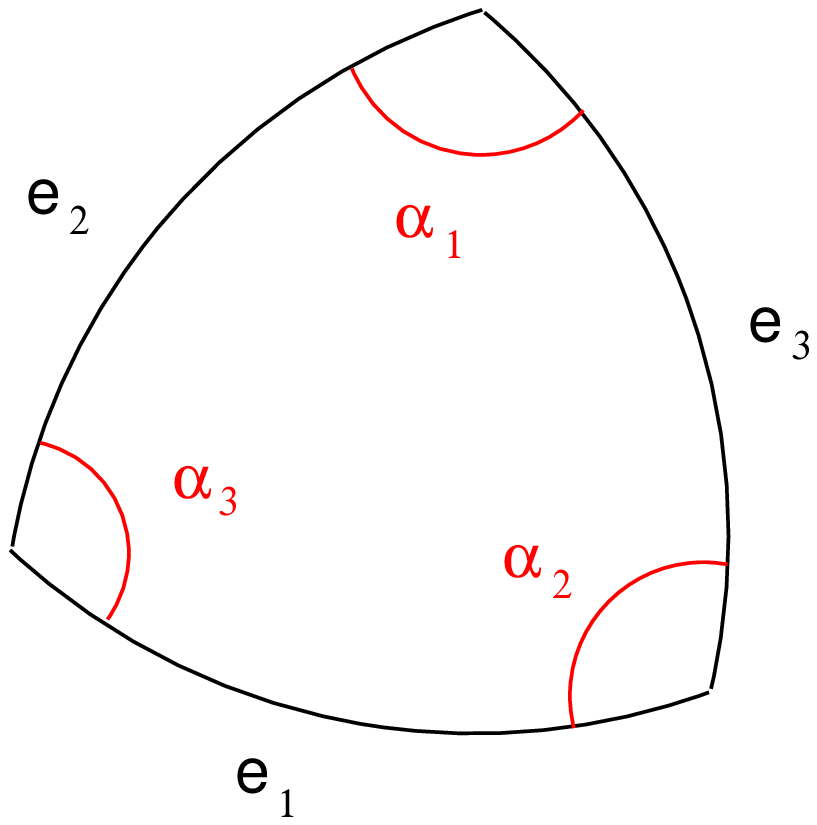}
  \small{{\bf figure 3b:} The elliptic triangle.}
\end{minipage}
\end{figure}
\fi

`t Hooft's vertex equations involve angles and the boost parameters 
of the Lorentz transformation. The equations of motion tell us how the 
lengths of the edges change with time. Nevertheless coordinates are not 
only necessary to set up a proper initial configuration but in addition 
if a vertex (later on also a particle) hits an edge and a transition takes 
place. To get the positions of the newly formed edges, coordinates and in 
particular a parametrization of the edges are needed.  
We continue to interpret the edge length in a Euclidean manner. 
Given two vertices $p_1, p_2$ the straight line $\overline{p_1 p_2}$ 
joining them is the length of the edge which is described by a circle 
segment through $p_1$ and $p_2$. 

If the angles $\ga_i$ add up to $2 \pi$, we are back in 
Euclidean geometry. Then the edges are straight lines. 

In this case we find two distinct configurations.  
Either $\eta_1 \: = \: \eta_2 \: = \: \eta_3 \: = \: 0$ or 
$\eta_2 \: = \: 0$ and 
$\eta_1 \: = \: \eta_3 \: = \: \eta$ (and permutations). 
The former case is known as ``trivial vertex'' in the literature 
\cite{FraGua95}. 
At a ``trivial vertex'' the angles $\alpha_1$, $\alpha_2$, and $\alpha_3$
can be chosen arbitrarily.
It is a special case of the latter one which is called the 
``quasistatic vertex'' (see \cite{FraGua95}). 
Let us demonstrate this with equation (1i): 
$\co_3 = \co_1 \co_2 \: - \: \si_1 \si_2 \ch_3$. 
For $\co_3$ we have $\co_3 = \cos(2 \pi - \ga_1 - \ga_2) 
= \co_1 \co_2 \: - \: \si_1 \si_2$. Therefore either $\si_1 \si_2 = 0$ or 
$\ch_3 = 1$.       
In the former case we get by a cyclic permutation of (1i) that either 
all of the velocities vanish 
or the vertex is a quasistatic one. 

\iffigs
\begin{figure}[bt]
\begin{center}
\begin{minipage}[t]{9cm}
  \epsfxsize=8cm\epsfbox{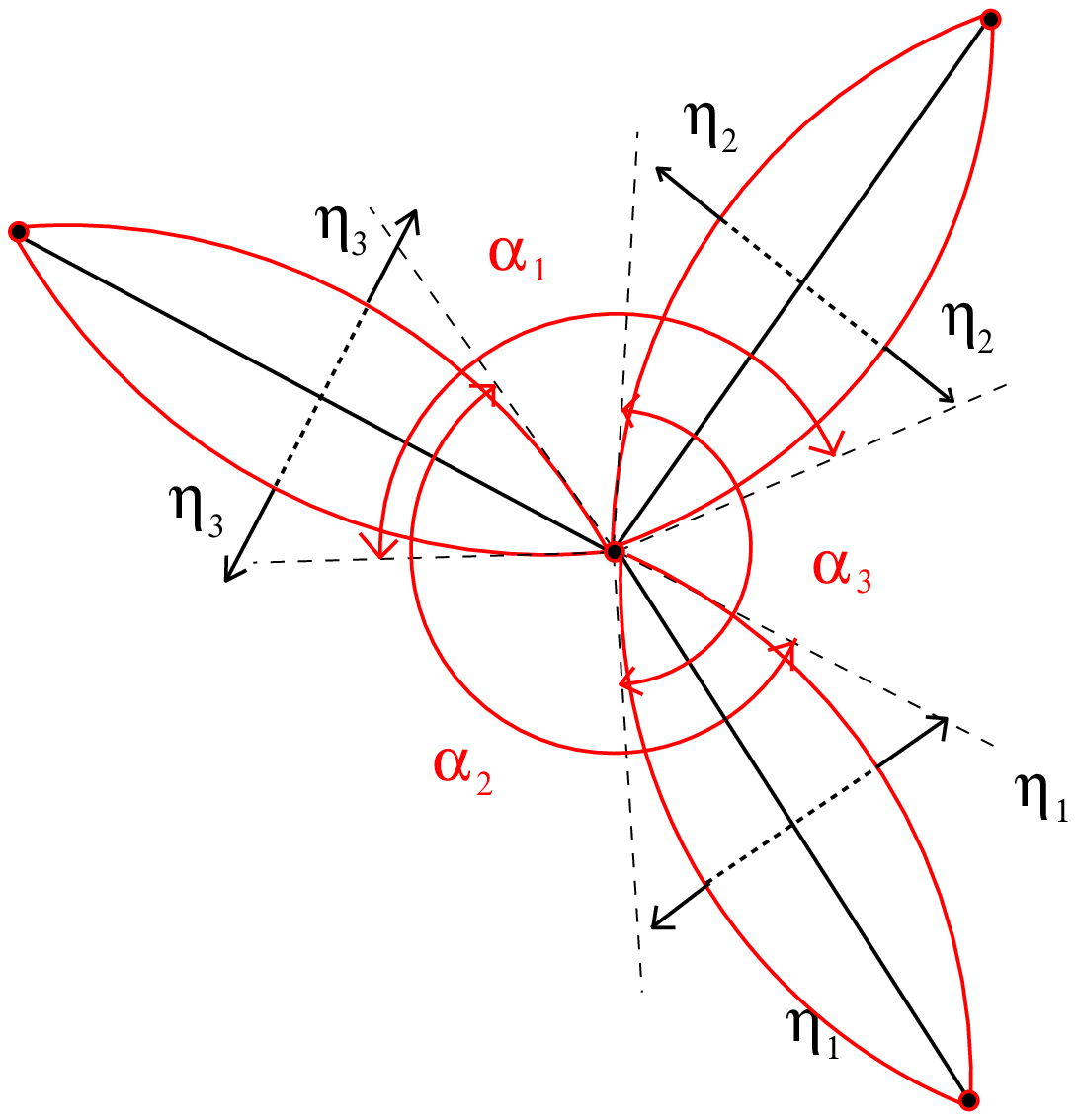}
  \small{{\bf figure 4:} The moving vertex}
\end{minipage}
\end{center}
\end{figure}
\fi

The latter case only contains the quasistatic vertex.
If one of the rapidities, say $\eta_1$, vanishes and the other two are  
equal the angles cannot be choosen freely. $\ga_1$ is equal to $\pi$, 
$\ga_2 = \ga$ and $\ga_3$ turns out to be $\pi - \ga$. 
 
The Euclidean configurations lead us to the interpretation of the 
non--Euclidean ones. Whenever we choose one of the $\eta$'s to be equal to  
zero the configuration is a Euclidean one. As nothing prevents us 
from doing so the hyperbolic configurations turn out to be pure 
gauge. They are not all of the possible gauge transformations as we 
initially fixed the gauge partially by setting $g_{00} = 1$. 
As an aside we remark that in order to extract physical 
information like the energy of the polygon universe we have to 
transform our tessellated surface to one common frame.

Let us make a Gedanken experiment. Assume that there is exactly one 
coordinate system attached to every polygon. Assume further that 
the coordinate systems of two polygons say {\bf I, II} (see {\bf figure 5}) 
transform into each other. We use the transformation procedure 
described above.  
Then the vertices cannot be treated in isolation in the polygon approach. 
The data which are connected with a vertex -- the angles and the 
boost velocities -- tell us how to perform a coordinate 
transformation from one polygon to another one across a common edge. 
Let us consider the situation shown in {\bf figure 5}. 
Let $O$ be the origin of frame I and II. 
We now perform the transformation across the edge $\overline{AB}$ 
in two ways. At first we go via $A$.  Let us assume that the coordinate 
axis makes an angle $\ga$ with the edge $\overline{AB}$. 
Another possibility is to go via $B$. 
Let us assume that the frame axes and $\overline{AB}$ there enclose  
an angle $\gb$. The two ways of performing the transformation 
of points $x$ in frame I to points $x^\prime$ in frame II must 
yield the same result.

\iffigs
\begin{figure}[bt]
\begin{center}
\begin{minipage}[t]{8cm}
  \epsfxsize=7cm\epsfbox{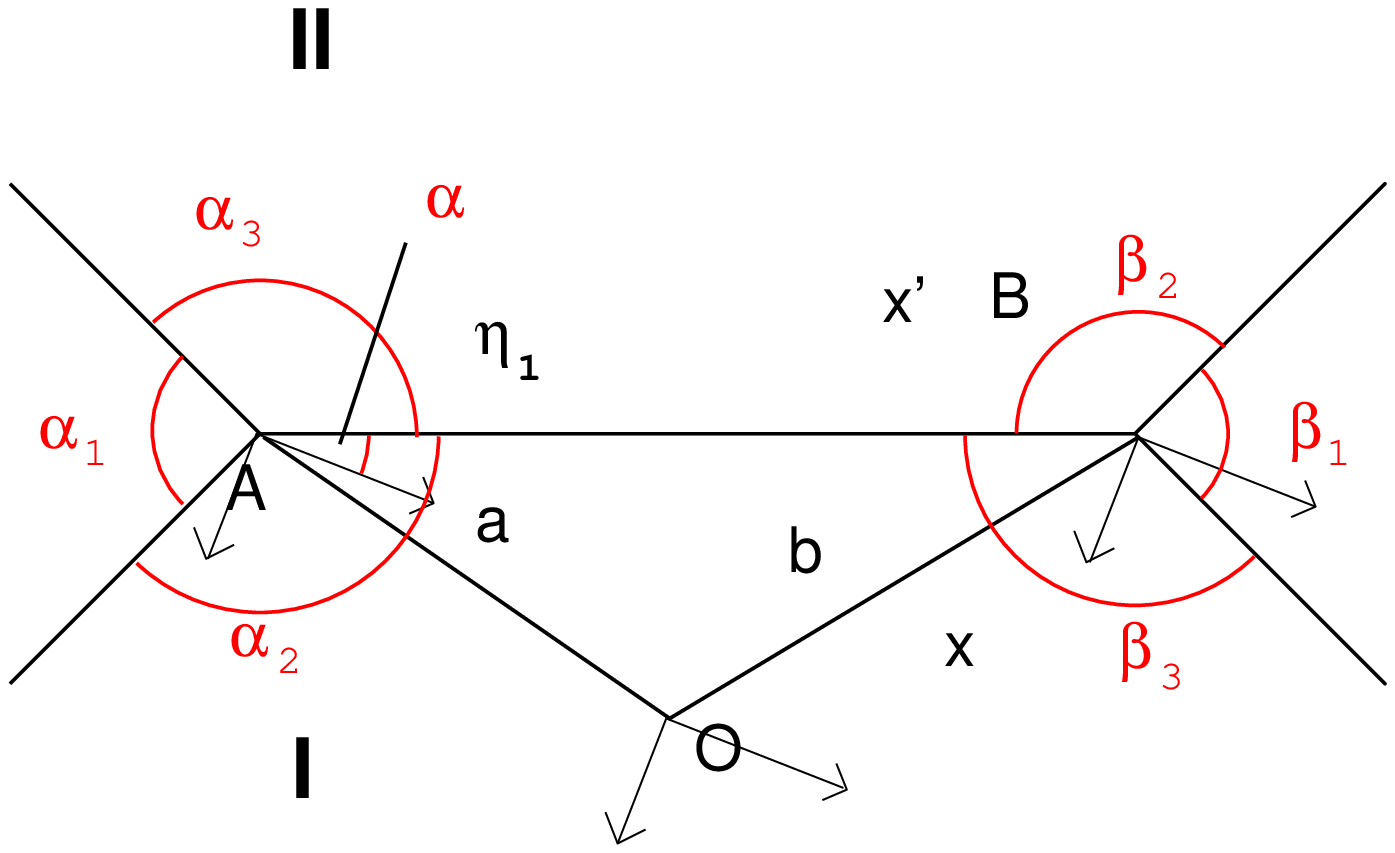}
  \small{{\bf figure 5:} Matching of vertices} 
\end{minipage}
\end{center}
\end{figure}
\fi

Let us start with a frame attached to $A$ denoted by the ``$A$--frame'' 
in the following. A transformation from I to II means rotating 
by an angle $\ga_3$ and applying a boost with parameter $2 \eta_1$:
\[
  x^\prime = R_{-\ga} R_{\ga_3} L_1 R_\ga  \: x.
\] 
If we attach the same frame to $B$ (called the $B$--frame) it 
follows that 
\[
  \overline{x}^\prime = R_{-\ga} R_{\gb_2} L_1 R_\ga  \: \overline{x}.
\] 
Points $x$ measured in the $A$--frame have to be transformed into 
points $\overline{x}$ in the $B$--frame and vice versa 
\[
  \overline{x} \:=\: x + a - b, \qquad 
  x \:=\:  \overline{x} +  b - a,  
\]
where $a$ and $b$ denote the vectors $\overline{OA}$ and 
$\overline{OB}$, respectively. 
We now describe the movements in the $B$--frame as seen with respect 
to the $A$--frame. 
Take a point $x$ in the $A$--frame, transform it first to a point 
$\overline{x}$ in the $B$--frame and then to $\overline{x}^\prime$ in 
frame II. Finally we take $x^\prime$ by the inverse transformation to 
the $A$--frame. 
The result is
\bea
  x^\prime &=&  R_{-\ga} R_{\gb_2} L_1 R_\ga  
    \: (x + a -b) + b-a  \nonumber \\
\eea  
On the way across $A$ we find 
\[
  x^\prime \:=\: R_{-\ga} R_{\ga_3} L_1 R_\ga \: x.
\]
Therefore we read off 
\bsea
   R_{-\ga} R_{\ga_3} L_1 R_\ga & = & R_{-\ga} R_{\gb_2} L_1 R_\ga \\
   R_{-\ga} R_{\gb_2} L_1 R_\ga \: (a-b) & = & a- b.
\esea
The first equation gives $\ga_3 = \gb_2$. 
The second equation tells that 
the edge $\overline{AB}$ is an eigenvector of the Lorentz transformation 
$R_{\gb_2} L_1$ with eigenvalue 1. 
This is nothing but the information that all points on this edge are 
fixed points under the transformation. 
That is, given the transformation law and given that there is one and 
only one frame attached to every polygon we get the result that  
't Hooft's polygon approach only allows for regular polygons in the 
pure gravity case. All angles in a polygon have to be equal. 
In particular that would mean given the boosts at any node in any 
polygon, all the angles and all the boosts in the whole universe 
are given by the vertex equations.  
As an aside we also like to comment on the quasistatic vertex. 
There one angle is forced to be equal to $\pi$. By the result above 
all angles within the polygon would have to be equal to $\pi$. 
As our polygons have to close we get the result that the quasistatic 
vertex in the pure gravity case would be a possible configuration in 
closed universes only, but not in open ones.  

We conclude that it is too restrictive to assume that there is one 
and only one coordinate system attached to every polygon. Depending 
on which vertex gets involved in the transformation from one polygon 
to an adjacent one we have to allow for frames which are rotated  
relative to each other.

\section{Tessellated Particle Universes in (2+1)D}

We now couple point particles to (2+1)D gravity. The resulting
equations of motion are exactly solvable and lead to a conical
structure of space--time. The angle of the cone is proportional
to the mass of the particle which is restricted by the equations
of motion to $0 \leq m \leq 1/4G$ in the one particle 
case\footnote{whenever the mass exceedes this limit the high mass 
particle sitting at the origin can be transformed into one with 
a mass within the range $0 \leq m \leq 1/4G$ sitting at $\infty$.} 
Many--body sources can have higher mass \cite{DesJac'tH84}.

\iffigs
\begin{figure}[bt]
\begin{center}
\begin{minipage}[t]{11cm}
  \epsfxsize=10cm\epsfbox{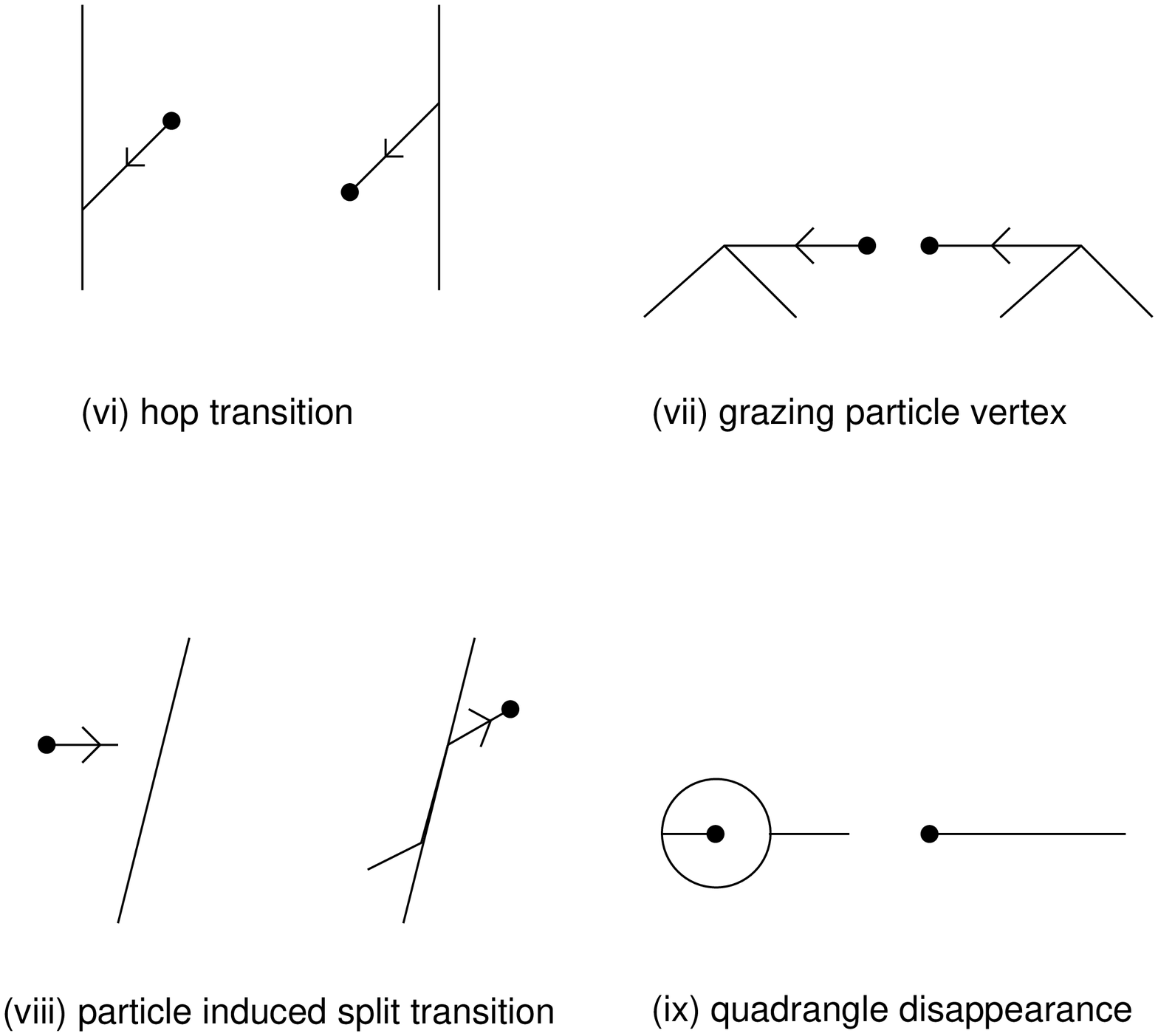}
  \vspace*{0.5cm}
  \small{{\bf figure 6:} particle transitions} 
\end{minipage}
\end{center}
\end{figure}
\fi

If we project the gravitational field of a point particle
onto the 2D Cauchy surface, the particle cuts out a wedge of the
2D hypersurface. 
The two sides of this wedge are identified.
As the space--time between the
particles still remains flat all particles have to be at vertex
points.
Besides the vacuum transitions so called particle transitions have
to be taken into account. They have been classified by 't Hooft
\cite{'tH92} and are shown in {\bf figure 6}.

If we consider an edge ending in a particle, i.e. a 1--valent particle 
vertex, the vertex equations get replaced by the particle equations. 
These particle consistency equations also have been introduced
in the paper \cite{'tH92}. They are derived by geometrical
considerations (see {\bf figure 7}). Let us denote by $ 2 \mu $ the 
angle of the wedge cut by the static particle in the 2D spatial 
hypersurface. Now we assume that the particle is moving with a velocity
$v = \tanh \xi$, where $\xi$ is the Lorentz boost parameter, in the
direction of the bisector of the wedge. Due to the motion of the particle 
the wedge gets a velocity $\tanh \eta$ as well.
By Lorentz contraction the wedge widens. We denote the angle by $2 H$ 
and read from {\bf figure 7} (ii)

\iffigs
\begin{figure}[bt]
\begin{center}
\begin{minipage}[t]{10cm}
  \epsfxsize=9cm\epsfbox{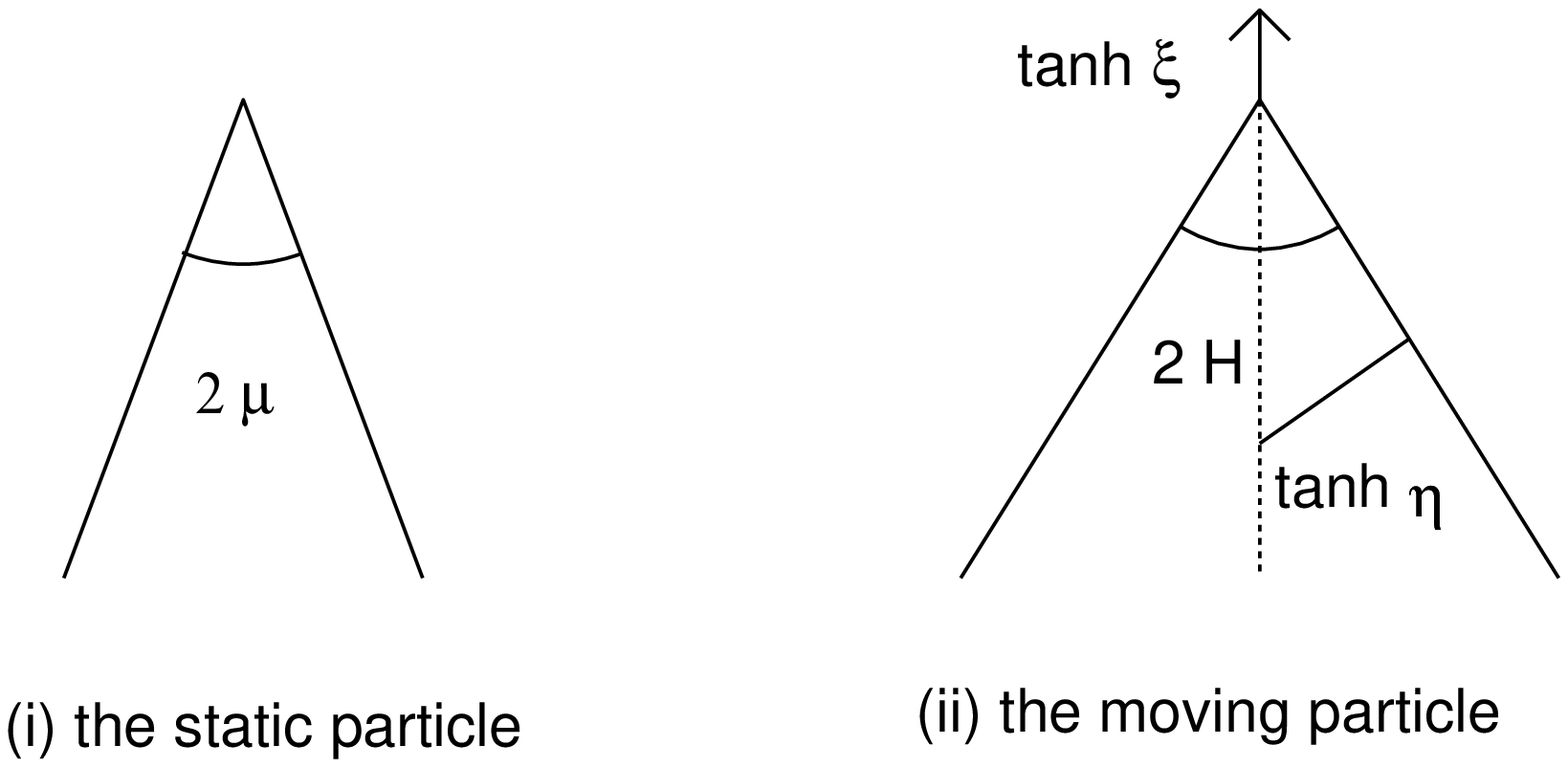}
  \small{{\bf figure 7:} particle equations} 
\end{minipage}
\end{center}
\end{figure}
\fi

\be
\label{pe1}
  \tanh \xi \sin H \:=\: \tanh \eta.
\ee
Combination of {\bf figure 7} (i) and (ii) yields 
\be
\label{pe2}
  \tan H \:=\: \tan \mu \cosh \xi.
\ee
From (\ref{pe1}) and (\ref{pe2}) the equations
\bea
  \label{pe3}
  \cos \mu &=& \cos H \cosh \eta \\
  \label{pe4}
  \sinh \eta &=& \sin \mu \sinh \xi
\eea
can be derived. 
Equation (\ref{pe1}) relates the boost velocity $\tanh \eta$ and the 
velocity of a vertex $\tanh \xi$. $H$ is half the deficit angle. 
That means whenever there is a boost neither the deficit angle nor the 
velocity of the vertex can be zero. If one assigns a mass to a vertex, 
a deficit angle at rest can be obtained by equation (\ref{pe3}). 
These equations are used to calculate proper initial
configurations and new sets of data after a particle involved
transition takes place.

Let us now define an angle $\hat{H}$ by setting $ H = \pi/2 + \hat{H}$. 
This implies $\sin H \:=\: $ \\ 
$\sin \pi/2 \: \cos \hat{H} \:=\: \cos \hat{H}$. 
In terms of the new quantity $\hat{H}$ equation (\ref{pe1}) becomes
\be
  \label{pe1p}
  \tanh \eta \:=\: \cos \hat{H} \tanh \xi
\ee
This equation and equation (\ref{pe4}) are nothing but the defining 
equations for a rectangular triangle in hyperbolic space, i.e. 
$\hat{H}, \mu, \xi$ may be interpreted as angles and edges in a rectangular 
triangle in hyperbolic geometry. 
This is illustrated in {\bf figure 8}. Two entities completely determine 
the triangle. Hence either $\eta$ and $\hat{H}$ or $\eta$ and $\xi$ can  
be chosen.

The particle could also be located at a 3--valent vertex, leading to 
different consistency equations. 
The presence of a particle curves the spacetime. That is, the vertex 
with its three joining edges is no longer in a flat 2D spatial hypersurface 
but is located on a cone. 
Let us first consider the case where the particle is at rest as 
illustrated in {\bf figure 9}. 
The points $A$ and $A^\prime$ are identified by a linear transformation. 
That means in this context that the sines of two angles $\gt$ and 
$\gt^\prime$ are equal if $\gt$ and $\gt^\prime$ differ by multiples 
of $2 (\pi - \mu)$. 
From this one can calculate the transformation from an angle 
$\ga^\prime$ in the plane to an angle $\ga$ on the cone, 
$\ga^\prime = c \ga$.  $c \in \R $ is the set of all linear transformations 
\bea 
  & & \sin 2 c (\pi - \mu) = 1 \\
  & \Longrightarrow & 2 c (\pi - \mu)  = 2 \pi k, \quad k \in \Z \\
  & \Longrightarrow & c = \frac{\pi}{\pi - \mu} \: k, \quad k \in \Z
\eea
If at a point $P$ with mass $m$ there meet three vertices with angles 
$\ga_1, \ga_2, \ga_3$, where $\ga_i$ denotes the angle in a reference 
frame of an origin with mass zero the new consistency conditions are 
given by 
\be
\label{parteqn1}
  L_2 \: R_3^\prime \: L_1 \: R_2^\prime \: L_3 \: R_1^\prime \: = \: \1,
\ee
where $R_i^\prime$ denotes a rotation by 
$\ga_i^\prime = \frac{\pi}{\pi - \mu} \: \ga_i$. In the limit 
$\mu \longrightarrow 0$ we find 
$\lim_{\mu \longrightarrow 0} \ga_i^\prime  = \ga_i$. The boosts $\eta_i$ 
of the edges are not changed. 

Let us now consider the case of a moving vertex with mass $m > 0$. 
There is a boost $\eta$ and equation (\ref{pe2}) holds. 
The angle $\mu$ is boosted by the velocity $\tanh \eta$ to become 
the angle $H$. 
This gives  
\be
\label{parteqn2}
  L_2 \: R_3^\prime \: L_1 \: R_2^\prime \: L_3 \: R_1^\prime \: = \: L_\eta,
\ee
In the limit of vanishing velocity we have $\eta \longrightarrow 0$ 
and one obtains the equation (\ref{parteqn1}) again. 
In the limit of vanishing mass the angle $\mu$ is zero as well and 
we are back to the situation without a particle, i.e. the massless 
vertex. Note that (\ref{parteqn2}) is not very different from the 
case where the particle is at rest. The boost $L_2$ is reduced by $\eta$: 
\[
  L_\eta^{-1} \: L_2 \: R_3^\prime \: L_1 \: R_2^\prime \: L_3 
    \: R_1^\prime \: = \: \1.
\]
As this equation holds for all cyclic permutations of the indices 
$1, 2, 3$ the assertion is true for the boosts $L_1$ and $L_3$ as well
\bea
  L_\eta^{-1} \: L_3 \: R_1^\prime \: L_2 \: R_3^\prime 
    \: L_1 \: R_2^\prime \:  &=& \1. \nonumber \\
  L_\eta^{-1} \: L_1 \: R_2^\prime \: L_3 \: R_1^\prime 
    \: L_2 \: R_3^\prime \:  &=& \1. \nonumber 
\eea
Using these three equations to solve for $\eta_1, \eta_2$ and $\eta_3$ 
in terms of the angles shows that the boost due to the particle is no 
longer observed at 3--valent vertices (although of course its effect 
is seen in the angle $H$).

The equations of motion of the polygon tessellated universe have to
be supplemented by another growth rate. An edge connects two vertices at
which there is either a particle or there is none. The growth rate
associated with the moving particle can be calculated as follows 
({\bf figure 10}).

\iffigs
\begin{figure}[bt]
\begin{center}
\begin{minipage}[t]{8cm}
  \epsfxsize=7cm\epsfbox{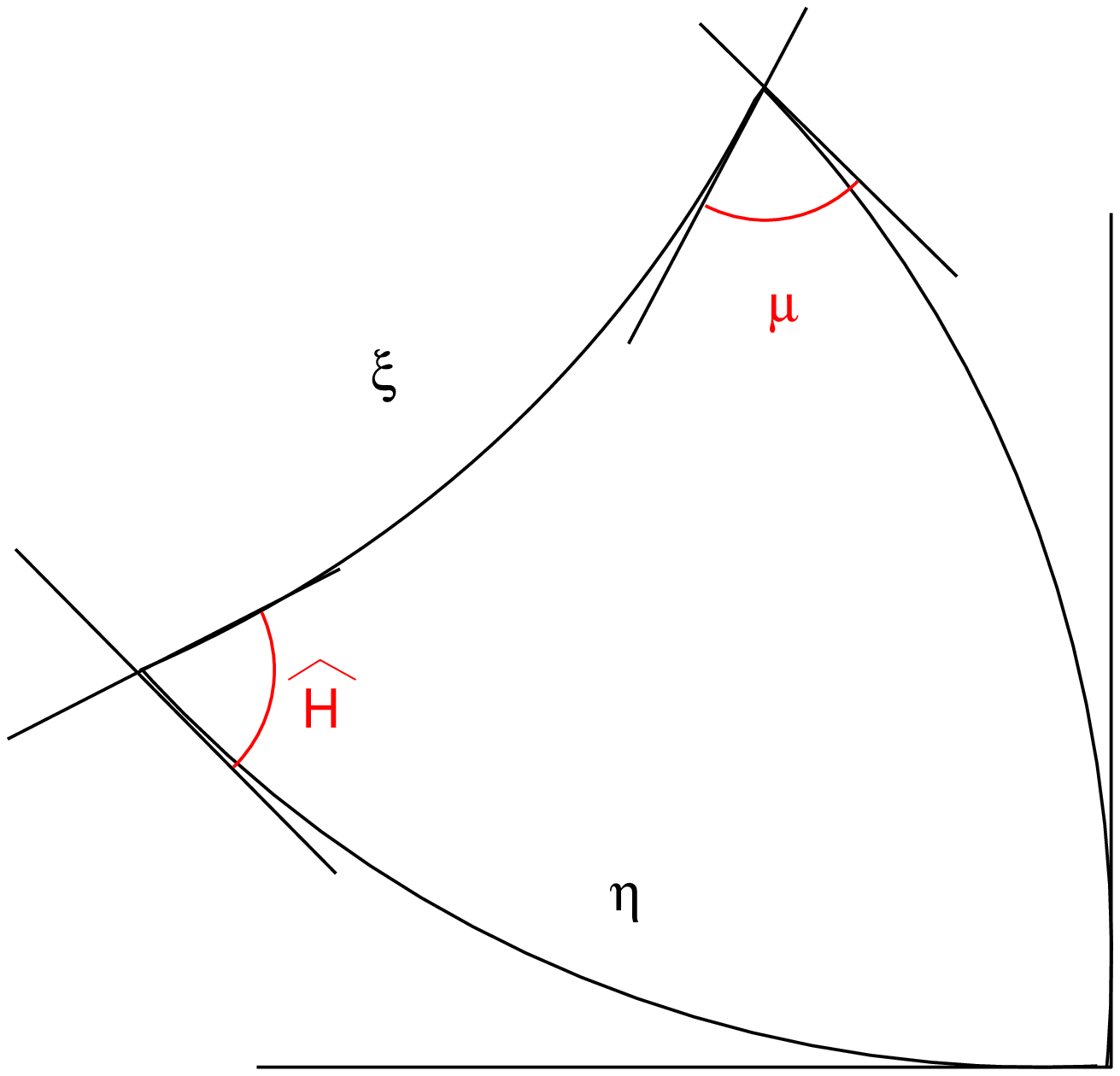}
  \small{{\bf figure 8:} hyperbolic geometry in the particle context} 
\end{minipage}
\end{center}
\end{figure}
\fi

\iffigs
\begin{figure}[bt]
\begin{center}
\begin{minipage}[t]{8cm}
  \epsfxsize=7cm\epsfbox{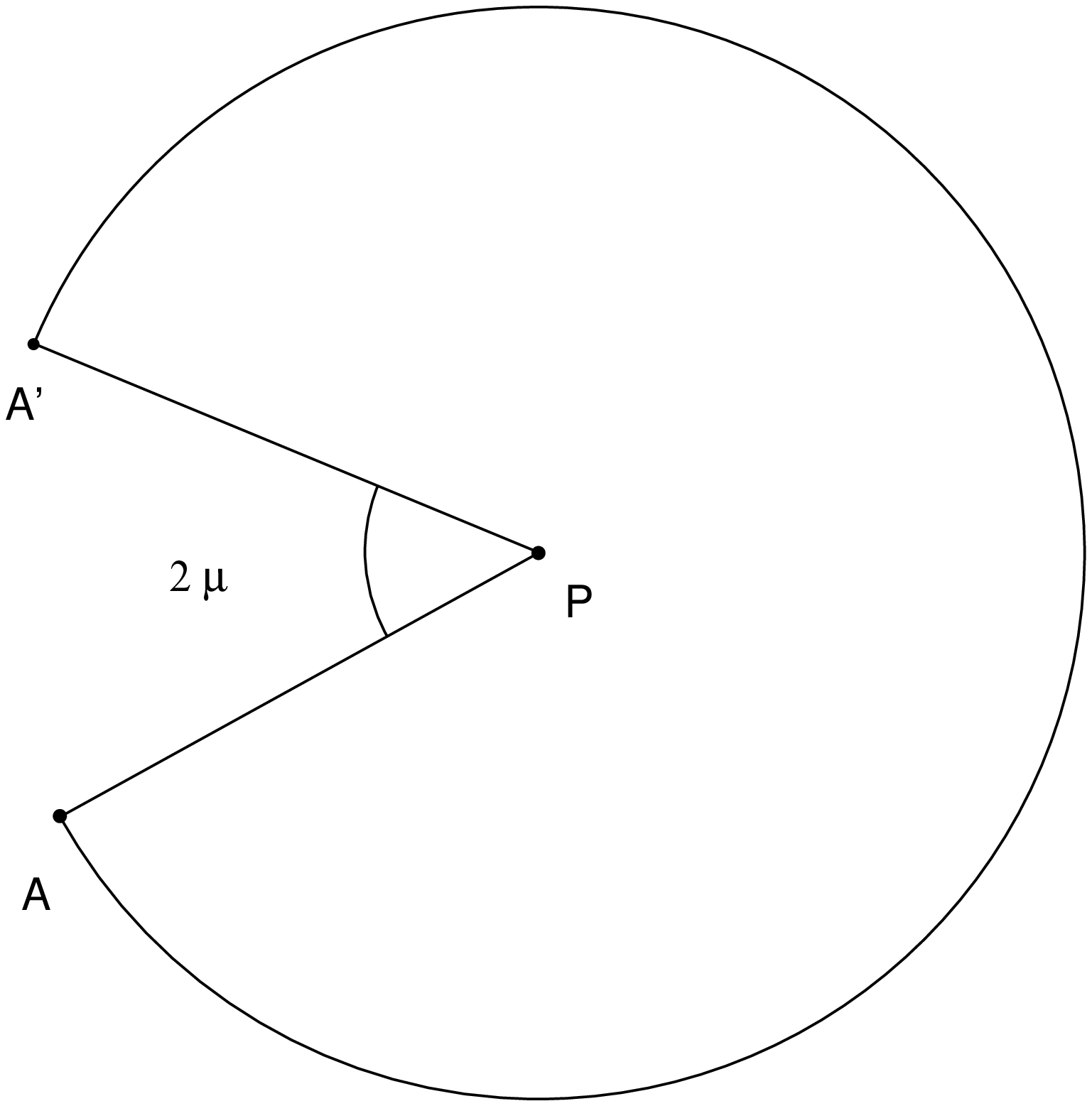}
  \small{{\bf figure 9:}  particle vertex at rest} 
\end{minipage}
\end{center}
\end{figure}
\fi

Assume a particle with mass $m$ is moving with constant velocity  
in the negative $x$--direction. Due to the mass of the particle and 
the motion it cuts out a wedge of the 2D spatial hypersurface with 
an angle $2H$ at a given time $t$.  At another instant of time , say 
$t + dt$, the wedge is translated in the direction of motion by an 
amount $\tanh \xi dt$. The growth rate is then 
\[
   g_P \:=\: \tanh \eta \cot H \:=\: \tanh \xi \cos H.
\]

\iffigs
\begin{figure}[bt]
\begin{center}
\begin{minipage}[t]{8cm}
  \epsfxsize=7cm\epsfbox{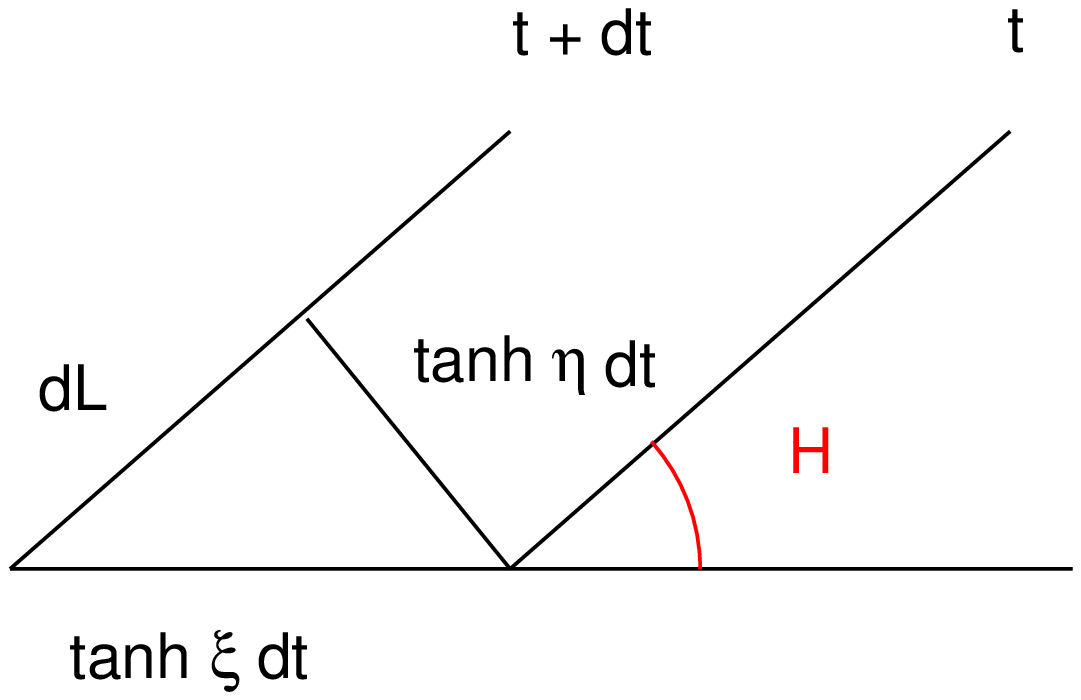}
\small{{\bf figure 10:} particle equations of motion} 
\end{minipage}
\end{center}
\end{figure}
\fi

\section{Some Initial Configurations}

At first we confine ourselves to pure gravity initial configurations 
and in particular to regular ones only. 
The consistency equations restrict the possible sets of data severely. 

In the following illustrations we have shown the Euclidean edge lengths 
and angles instead of the hyperbolic shapes only. All figures represent 
3D geometries.  {\bf Figure 11} shows a tetrahedral configuration drawn 
from the top. All boosts and all 
angles are equal. In {\bf figure 12} the configuration is cubical. 
The rectangles on the top and bottom are of equal size. In the figure one 
of them is drawn smaller in order to show the vertical edges. Neither 
the angles nor the edges need to be of equal size. 
{\bf Figure 13} and {\bf figure 14} differ from {\bf figure 12} by the shape 
of the top and the bottom polygon. In {\bf figure 13} the base polygon 
is a triangle and in {\bf figure 14} it is a pentagon. The restrictions on the 
angles and the boost parameters are found in the illustration. 
Thus we have introduced initial configurations with an even and an odd number 
of edges. 

Now we like to present initial data for regular and distorted particle 
universes. We place at every vertex $A, B, C, D$ of the tetrahedron 
(see {\bf figure 15}) a particle with mass $m_i$ and velocity 
$v_i, \: i = A, B, C, D$.  
The deficit angles $H_A, H_B, H_C$ and $H_D$ are then given by 
(\ref{pe2}), where $\mu_i = \pi \: m_i$ and 
$\cosh \xi = ( 1 - v_i^2/c^2 )^{-1/2}$. $c$ denotes the velocity of light. 
At the vertex $A$ we choose all angles $\ga_{A,j}, j = 1, 2, 3$  on the 
cone to be equal and calculate the angles $\ga_j^\prime$. 
$\ga^\prime_{A,j}$ are the angles which correspond to $\ga_{A,j}$ in 
the massless situation, i.e. 
$\ga_{A,j}^\prime = \frac{\pi}{\pi - H_A} \: \ga_{A,j}$. The boosts $\eta_{AB}, 
\eta_{AC}$ and $\eta_{AD}$ follow from equation (A.4) where the angles 
$\ga_j$ are replaced by the angles $\ga_j^\prime$. 
At vertex $B$ the boost $\eta_{AB}$ is already known. The angles 
$\ga_{CBA}, \ga_{ABD}$ are chosen arbitrarily, the corresponding angles 
$\ga_{CBA}^\prime, \ga_{ABD}^\prime$ are calculated.  The angle  
$\ga^{\prime}_{DBC}$ then the boosts $\eta_{BC}$ and $\eta_{BD}$ follow 
using the modified equation (A.4). 
At $C$ the boosts $\eta_{AC}$ and $\eta_{BC}$ are fixed. We can only 
give data for the angle $\ga_{ACB}$. $\eta_{CD}, \ga_{DCB}^\prime$ and    
$\ga_{DCA}^\prime$ follow from the modified equation (A.5).  
At $D$ all boosts are known and all angles have to be computed from (A.5). 

The numerical data for one of these distorted 4--particle tetrahedron 
configurations is included in {\bf table 1}.  

The completely regular tetrahedron is a special case of the situation 
described above. Setting the masses of the particles $m = 1/4$, the 
velocities $v = \sqrt{3}/2 \: c$ and the angles on the cone 
$\ga = 90^\circ$, one finds the deficit angles $H = 63.43^\circ$, 
the angles  $\ga^\prime = 138.98^\circ$, and the boosts  $\eta = 0.89$.  

{\footnotesize \begin{tabular}{l||l|l|l|l} \hline
& & & & \\
Vertex & A & B & C & D \\ \hline \hline
& & & & \\  
Mass & $m_A = 0.1$ & $m_B = 0.15$ & $m_C = 0.2$ & $m_D = 0.25$  \\ \hline 
& & & & \\
Velocity & $v_A = 0.5 \: c$ & $v_B = 0.6 \: c$ & 
    $v_C = 0.4 \: c$ & $v_D = 0.45 \: c$ \\  \hline 
& & & & \\
Deficit angle & $H_A = 20.57$ & $H_B = 32.49$ &  
    $H_C = 38.40$  & $H_D = 48.23$ \\  \hline 
& & & & \\
Observed & $\ga_{CAD} = 115.00$ & 
    $\ga_{DBC} = 111.59$ & $\ga_{BCD} = 94.02$ & 
    $\ga_{CDB} = 87.50$  \\
angles & $\ga_{DAB} = 115.00$ & $\ga_{CBA} = 105.00$ & 
        $\ga_{DCA} = 77.65$ & $\ga_{BDA} = 116.32$  \\
 & $\ga_{BAC} = 115.00$ & $\ga_{ABD} = 105.00$ & 
        $\ga_{ACB} = 125.00$ & $\ga_{ADC} = 72.26$  \\ \hline 
& & & & \\
Angles for & $\ga_{CAD}^\prime = 129.83$ & 
    $\ga_{DBC}^\prime = 136.18$ & $\ga_{BCD}^\prime = 119.53$ & 
    $\ga_{CDB}^\prime = 119.53$  \\
calculation & $\ga_{DAB}^\prime = 129.83$ & $\ga_{CBA}^\prime = 128.13$ & 
        $\ga_{DCA}^\prime = 98.71$ & $\ga_{BDA}^\prime = 158.90$  
   \\
 & $\ga_{BAC}^\prime = 129.83$ & $\ga_{ABD}^\prime = 128.13$ & 
        $\ga_{ACB}^\prime = 158.90$ & $\ga_{ADC}^\prime = 98.71$  
   \\ \hline 
& & & & \\
Boosts & $\eta_{AB} = 0.59$ & $\eta_{AB} = 0.59$ & 
       $\eta_{AC} = 0.59$ & $\eta_{AD} = 0.59$   \\ 
       & $\eta_{AC} = 0.59$ & $\eta_{BC} = 0.64$ &  
       $\eta_{BC} = 0.64$ & $\eta_{BD} = 0.64$  \\
       & $\eta_{AD} = 0.59$ & $\eta_{BD} = 0.64$   & 
       $\eta_{CD} = 0.29$  &  $\eta_{CD} = 0.29$   \\  \hline 
\end{tabular}}

\vspace*{0.5cm}

{\small {\bf table 1 :} Data for a distorted 4--particle tetrahedron 
universe }

\iffigs
\begin{figure}[bt]
\begin{center}
\begin{minipage}[t]{8cm}
  \epsfxsize=7cm\epsfbox{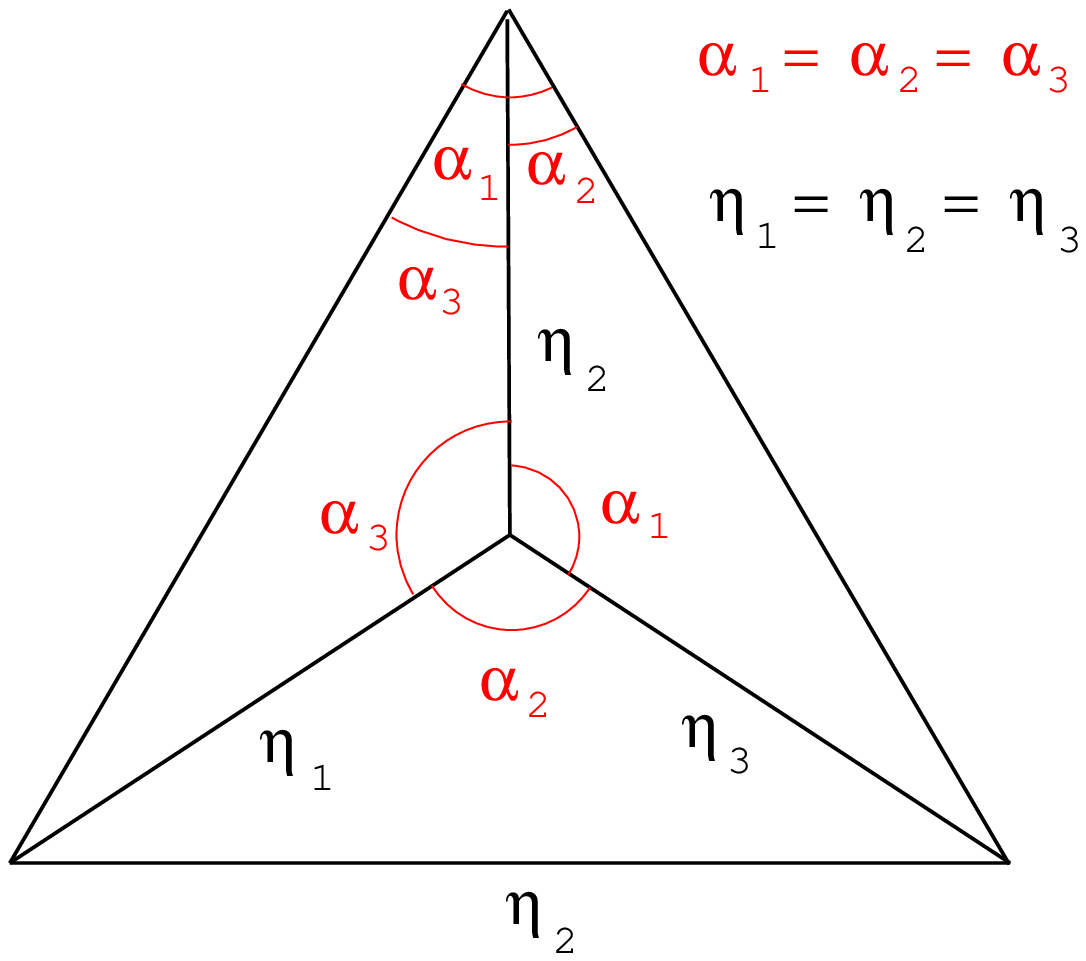}
  \small{{\bf figure 11:} initial configuration I} 
\end{minipage}
\end{center}
\end{figure}
\fi 


\section*{Acknowledgment}\addcontentsline{toc}
{section}{Acknowledgment}

One of the authors (H. H.) is indebted to G. 't Hooft, W. Glei{\ss}ner, 
R. Ionicioiu, and M. Welling for discussions on the subject.
The work has been supported in part by the UK Particle Physics and 
Astronomy Research Council. 
 
\newpage
\clearpage

\iffigs
\begin{figure}[bt]
\begin{center}
\begin{minipage}[t]{8cm}
  \epsfxsize=7cm\epsfbox{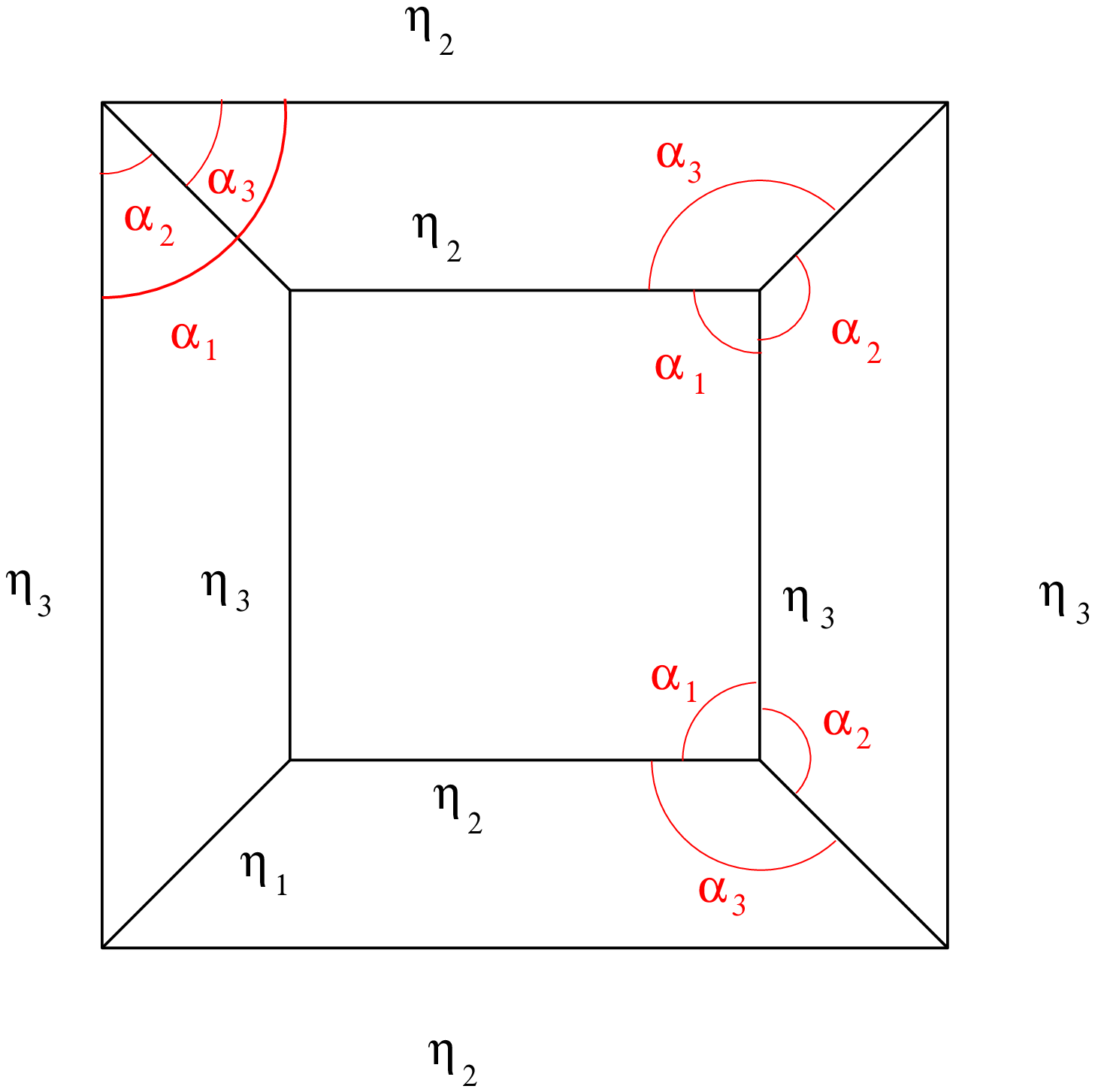}
  \small{{\bf figure 12:} initial configuration II} 
\end{minipage}
\end{center}
\vspace*{5cm}
\begin{center}
\begin{minipage}[t]{8cm}
  \epsfxsize=7cm\epsfbox{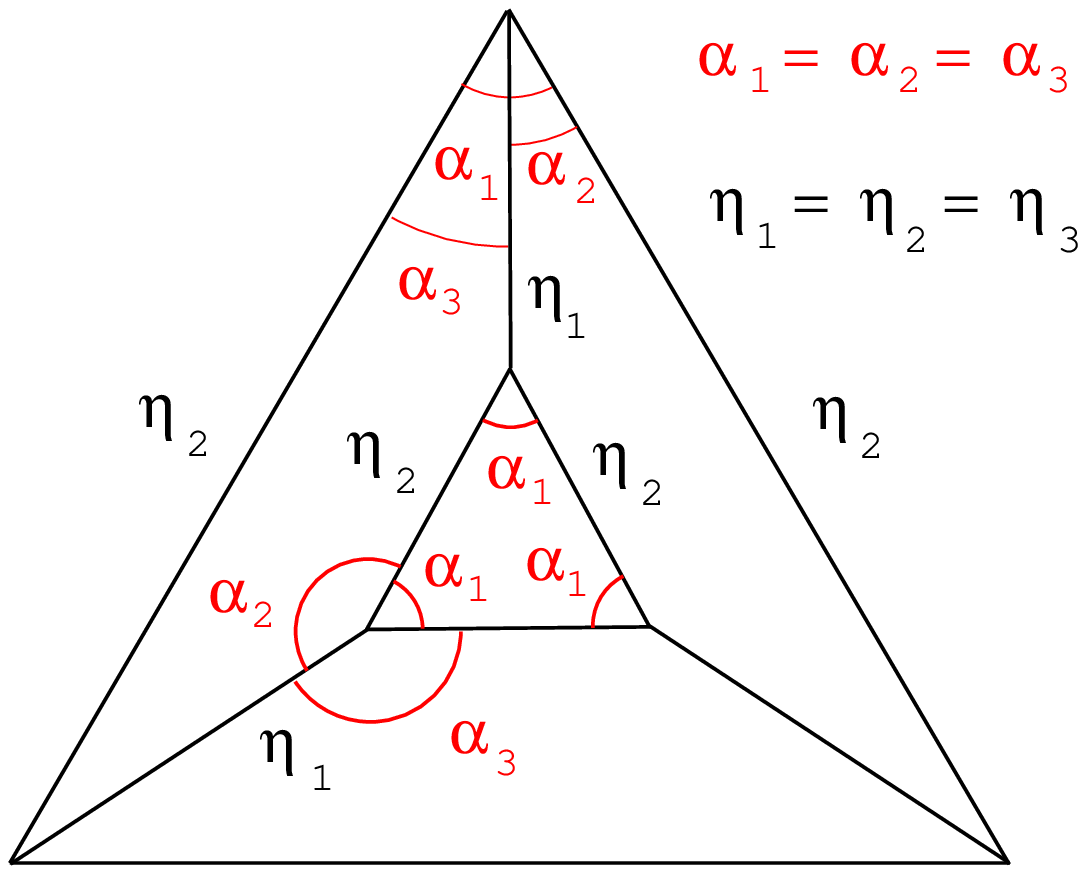}
  \small{{\bf figure 13:} initial configuration III} 
\end{minipage}
\end{center}
\end{figure}
\fi

\newpage
\clearpage

\iffigs
\begin{figure}[bt]
\begin{center}
\begin{minipage}[t]{8cm}
  \epsfxsize=7cm\epsfbox{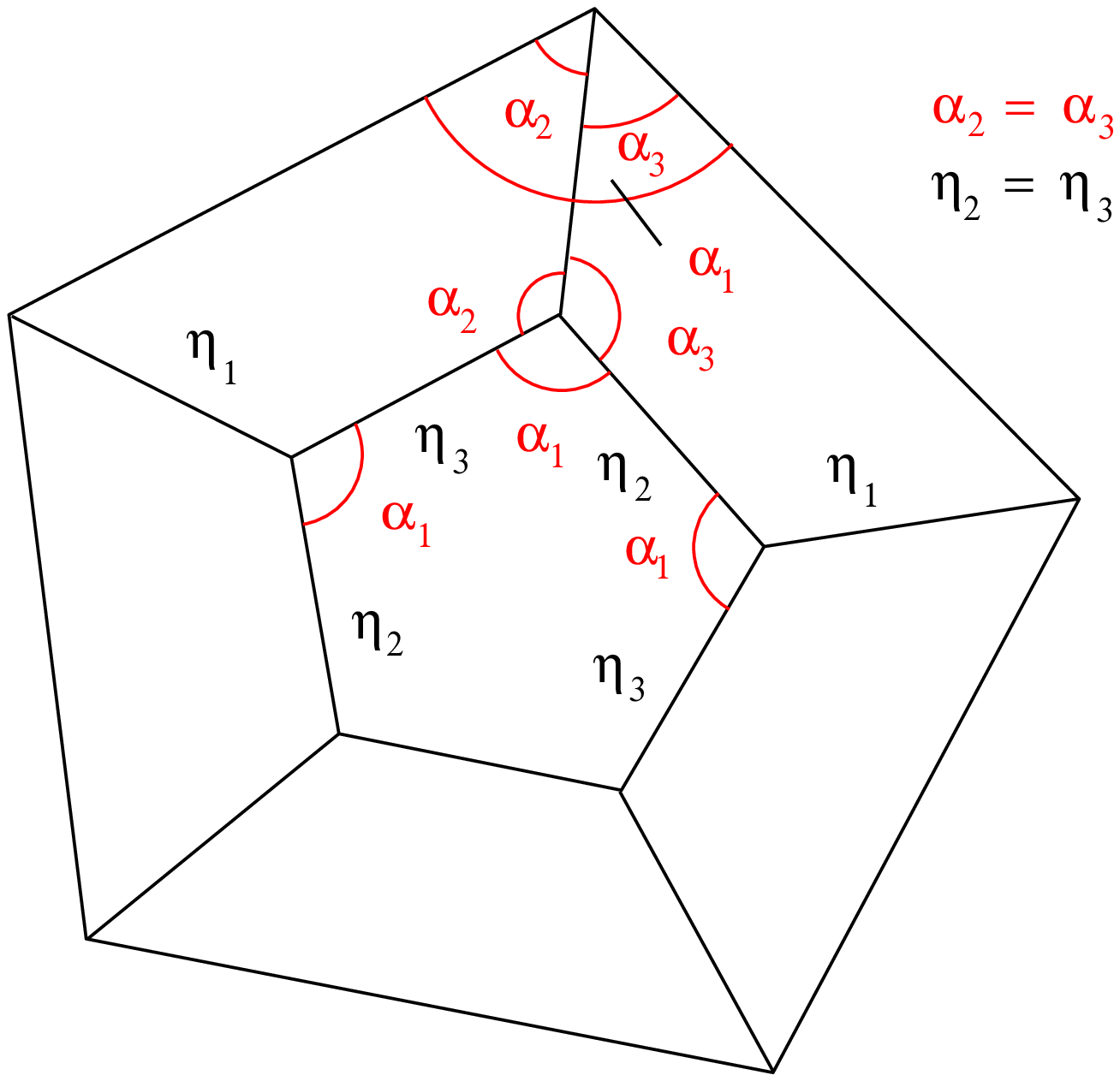}
  \small{{\bf figure 14:} initial configuration IV} 
\end{minipage}
\end{center}
\vspace*{5cm}
\begin{center}
\begin{minipage}[t]{8cm}
  \epsfxsize=7cm\epsfbox{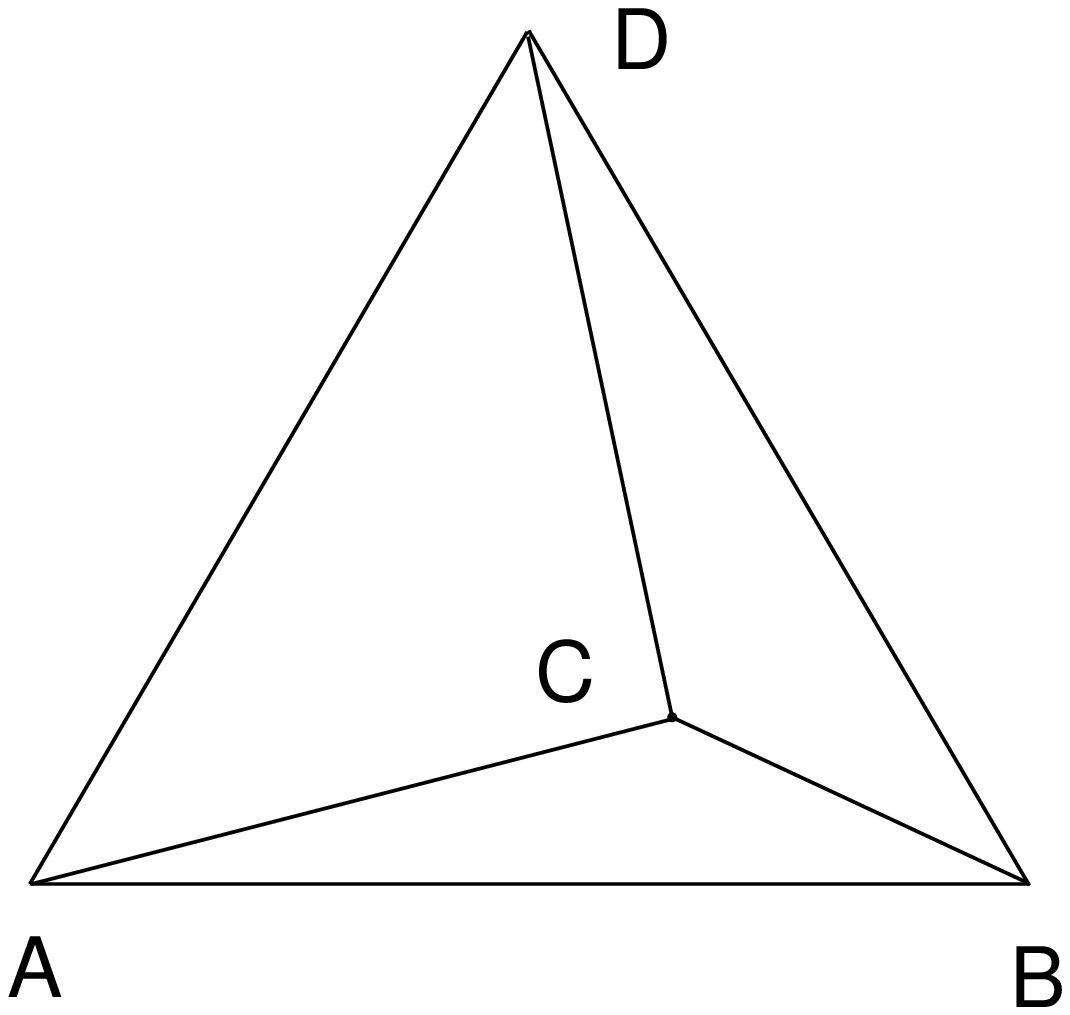}
  \small{{\bf figure 15:} distorted tetrahedron} 
\end{minipage}
\end{center}
\end{figure}
\fi


\end{document}